\documentclass[twocolumn]{aastex631}

\shorttitle{NIR Features of KNe: The Importance of Gd}
\shortauthors{Rahmouni et al.}

\begin{document}

\title{Revisiting Near-Infrared Features of Kilonovae: The Importance of Gadolinium}

\correspondingauthor{Salma Rahmouni}
\email{rahmouni.salma@astr.tohoku.ac.jp}

\author[0009-0002-1232-243X]{Salma Rahmouni}
\affiliation{Astronomical Institute, Tohoku University, Aoba, Sendai 980-8578, Japan}

\author[0000-0001-8253-6850]{Masaomi Tanaka}
\affiliation{Astronomical Institute, Tohoku University, Aoba, Sendai 980-8578, Japan}
\affiliation{Division for the Establishment of Frontier Sciences, Organization for Advanced Studies, Tohoku University, Sendai 980-8577, Japan}

\author[0000-0002-7415-7954]{Nanae Domoto}
\affiliation{Astronomical Institute, Tohoku University, Aoba, Sendai 980-8578, Japan}

\author[0000-0002-5302-073X]{Daiji Kato}
\affiliation{National Institute for Fusion Science, 322-6 Oroshi-cho, Toki 509-5292, Japan}
\affiliation{Interdisciplinary Graduate School of Engineering Sciences, Kyushu University, Kasuga, Fukuoka 816-8580, Japan}

\author[0000-0002-2502-3730]{Kenta Hotokezaka}
\affiliation{Research Center for the Early Universe, Graduate School of Science, University of Tokyo, Bunkyo, Tokyo 113-0033, Japan}

\author[0000-0002-8975-6829]{Wako Aoki}
\affiliation{National Astronomical Observatory of Japan, 2-21-1 Osawa, Mitaka, Tokyo 181-8588, Japan}
\affiliation{Astronomical Science Program, The Graduate University for Advanced Studies, SOKENDAI, 2-21-1 Osawa, Mitaka, Tokyo 181-8588,
Japan}

\author[0000-0003-3618-7535]{Teruyuki Hirano}
\affiliation{Astrobiology Center, 2-21-1 Osawa, Mitaka, Tokyo 181-8588, Japan}
\affiliation{National Astronomical Observatory of Japan, 2-21-1 Osawa, Mitaka, Tokyo 181-8588, Japan}
\affiliation{Astronomical Science Program, The Graduate University for Advanced Studies, SOKENDAI, 2-21-1 Osawa, Mitaka, Tokyo 181-8588,
Japan}

\author[0000-0001-6181-3142]{Takayuki Kotani}
\affiliation{Astrobiology Center, 2-21-1 Osawa, Mitaka, Tokyo 181-8588, Japan}
\affiliation{National Astronomical Observatory of Japan, 2-21-1 Osawa, Mitaka, Tokyo 181-8588, Japan}
\affiliation{Astronomical Science Program, The Graduate University for Advanced Studies, SOKENDAI, 2-21-1 Osawa, Mitaka, Tokyo 181-8588,
Japan}

\author[0000-0002-4677-9182]{Masayuki Kuzuhara}
\affiliation{Astrobiology Center, 2-21-1 Osawa, Mitaka, Tokyo 181-8588, Japan}
\affiliation{National Astronomical Observatory of Japan, 2-21-1 Osawa, Mitaka, Tokyo 181-8588, Japan}

\author[0000-0002-6510-0681]{Motohide Tamura}
\affiliation{Astrobiology Center, 2-21-1 Osawa, Mitaka, Tokyo 181-8588, Japan}
\affiliation{National Astronomical Observatory of Japan, 2-21-1 Osawa, Mitaka, Tokyo 181-8588, Japan}
\affiliation{Department of Astronomy, Graduate School of Science, The University of Tokyo, 7-3-1 Hongo, Bunkyo-ku, Tokyo 113-0033, Japan}

\begin{abstract}

The observation of the kilonova AT2017gfo and investigations of its light curves and spectra confirmed that neutron star mergers are sites of \(r\)-process nucleosynthesis. However, the identification of elements responsible for the spectral features is still challenging, particularly at the near-infrared wavelengths. In this study, we systematically searched for all possible near-infrared transitions of heavy elements using experimentally calibrated energy levels. Our analysis reveals that most candidate elements with strong absorption lines are lanthanides (\(Z=57-71\)) and actinides (\(Z=89-103\)). This is due to their complex structures leading to many low-lying energy levels, which results in strong transitions in the near-infrared range. \cite{domoto2022lanthanide} have shown that La III and Ce III can explain the absorption features at \(\lambda\sim\) \(12,000 - 15,000\,\text{\AA}\). While our results confirm that these two elements show strong infrared features, we additionally identify Gd III as the next most promising species. Due to its unique atomic structure involving the half-filled \(4f\) and the outer \(5d\) orbitals, Gd III has one of the lowest-lying energy levels, between which relatively strong transitions occur. We also find absorption lines caused by Gd III in the near-infrared spectrum of a chemically peculiar star HR 465, which supports their emergence in kilonova spectra. By performing radiative transfer simulations, we confirm that Gd III lines affect the feature at \(\sim 12,000\,\text{\AA}\) previously attributed to La III. Future space-based time-series observations of kilonova spectra will allow the identification of Gd III lines.
\end{abstract}


\section{Introduction}\label{sec:intro}

Binary neutron star mergers have long been considered a significant site for heavy element nucleosynthesis via the rapid neutron capture process, or \(r\)-process (e.g., \citealt{lattimer1974black,eichler1989nucleosynthesis, freiburghaus1999r, goriely2011r, wanajo2014production}). The electromagnetic emission from these events, known as kilonovae, is powered by the radioactive decay of the freshly synthesized nuclei \citep{li1998transient,metzger2010electromagnetic,roberts2011electromagnetic}. After gravitational waves were detected from a binary neutron star merger in 2017 (GW170817, \citealt{abbott2017gw170817}), an electromagnetic counterpart (AT2017gfo, \citealt{Abbott_2017}) was observed across infrared, optical, and ultraviolet wavelengths. The emission was consistent with the theoretical expectations of a kilonova (e.g., \citealt{arcavi2017optical,pian2017spectroscopic,smartt2017kilonova,utsumi2017j}), providing direct observational evidence of the occurrence of \(r\)-process in neutron star mergers (e.g., \citealt{kasen2017origin,shibata2017modeling,tanaka2017kilonova,kawaguchi2018radiative}).

Detailed spectra of AT2017gfo were obtained between 1.5 and 10 days after the merger (e.g., \citealt{pian2017spectroscopic, smartt2017kilonova, tanvir2017emergence}), spanning from ultraviolet to near-infrared (NIR) wavelengths. Initially, the spectra were dominated by photospheric emission with multiple absorption features. As the photosphere receded, AT2017gfo transitioned into the nebular phase around 7 days after the merger, revealing several atomic emission lines \citep{hotokezaka2021nebular,hotokezaka2022,hotokezaka2023,pognan2022a,pognan2022b,pognan2023nlte}.

It is necessary to uncover the abundance patterns produced by neutron star mergers to investigate whether they are the main site of \(r\)-process nucleosynthesis.
To achieve this goal, significant efforts have been made to identify the features observed in the kilonova photospheric spectra. \cite{watson2019identification} proposed that the spectral feature at \(9,000\,\text{\AA}\) could be explained by Sr II, a finding later confirmed with radiative transfer simulations \citep{domoto2021signatures}. Similar studies by \cite{gillanders2022modelling} verified Sr II impact and constrained its abundance. Alternatively, He I was suggested as a potential explanation for the same feature \citep{perego2022production}, with its significance being higher under non-local thermodynamic equilibrium (non-LTE) conditions \citep{tarumi2023non}. However, \cite{sneppen2024helium} recently highlighted some inconsistencies with the He I interpretation. The feature at \(7,000\,\text{\AA}\) was attributed to Y II by \cite{sneppen2023discovery} but Rb I was also considered, taking into account non-LTE effects \citep{pognan2023nlte}.

\cite{domoto2022lanthanide} tackled the investigation of the NIR absorption features of kilonovae. They demonstrated that elements belonging to the left side of the periodic table, such as Ca, Sr, Y, Zr, Ba, La, and Ce, are most likely to influence the kilonova spectrum. By performing radiative transfer simulations, they showed that La III and Ce III could account for the features at \(12,000\,\text{\AA}\) and \(14,000\,\text{\AA}\), respectively (see also \citealt{domoto2023transition}). In addition, \cite{tanaka2023cerium} showed that the strong absorption lines from Ce III appear in the NIR spectra of the chemically peculiar star HR 465, which matched the kilonova absorption features, further supporting previous findings.

The heaviest elements at the third \(r\)-process peak and beyond may also have been synthesized in GW170817 \citep{wanajo2014production}, but spectral features from such elements are yet to be identified. \cite{gillanders2021gold} confirmed that the third-peak elements, such as Au and Pt, do not produce detectable features. On the other hand, \cite{domoto2024} suggest that Th III is likely to exhibit features at \(\sim 18,000\,\text{\AA}\). However, this wavelength range overlaps with the region of atmospheric absorption, making the identification of Th III difficult in the case of AT2017gfo.

A thorough investigation of the spectral features remains challenging due to the lack of complete experimental atomic data. Significant theoretical efforts have been made to address this gap (e.g., \citealt{tanaka2020systematic, radzuite2020, radzuite2021, fontes2020, fontes2023, flors2023, gallego2023, gallego2024, kato2024}). Although these theoretical calculations are essential for opacity estimates, they do not provide the accuracy required for reliable spectral identifications. This lack of completeness and accuracy of atomic data suggests that unidentified transitions may still contribute to the observed absorption features. \cite{gillanders2024modelling} conducted a comprehensive search for important species to explain each feature in the spectra of AT2017gfo. However, they did not provide specific identifications. This necessitates further investigation to determine whether the spectral features, especially the least investigated NIR ones, could be attributed to other elements, and to give new identifications in the framework of the currently available atomic data.

In this paper, we narrow down the elements most likely to impact the kilonova spectra, aiming to discover new candidates for the observed spectral features. Our primary focus is on the NIR absorption features that appear during the early phase when the photospheric emission dominates. In Section \ref{sec:cand}, we use experimentally constructed energy levels to select elements with strong transitions in the NIR wavelengths. Section \ref{sec:gd} discusses the key properties of Gd III, identifying it as the most significant uninvestigated element among the selected candidates. We then perform radiative transfer simulations in Section \ref{sec:rad} to explore the effect of Gd III on the spectra and discuss our findings in Section \ref{sec:disc}. Finally, we give our conclusions in Section \ref{sec:conc}.

\section{Candidate Elements}\label{sec:cand}
\subsection{Method}\label{subsec:cand_method}

To find candidate elements that are likely to exhibit strong features in the NIR wavelength range, we started with the National Institute of Standards and Technology Atomic Spectra Database (NIST ASD, \citealt{nist}). This database contains experimental radiative transitions from which energy levels of elements are constructed. Since most spectroscopic experiments are conducted in the optical wavelength range, the infrared transition data available in the NIST ASD is largely incomplete. Thus, by using the energy differences between the constructed energy levels, we instead selected the allowed NIR transitions likely to occur and calculated the corresponding wavelengths. In this way, we constructed a new line list with infrared transitions of ions with atomic numbers \(30\leq Z\leq 90\). Due to a notable lack of energy level data for heavier elements in the NIST ASD, we used the Selected Constants Energy Levels and Atomic Spectra of Actinides (SCASA; \citealt{scasa}) to similarly select NIR transitions of the elements with \(91\leq Z\leq 99\). We note that we excluded all levels with uncertainties in their energies and/or their total angular momentum \(J\).
For Nd III (\(Z=60\)), we used the energy levels from \cite{ding2024spectrum} due to the completeness of their data.

We focused in our selection on the strong electric dipole transitions satisfying the parity change and the total angular momentum change (\(\Delta J = 0, \pm 1\) except \(0 \leftrightarrow 0\)) selection rules. It is important to note that we have not considered the selection rules of the \(LS\)-coupling scheme. Our focus in this paper is on heavy elements for which the spin-orbit interactions become strong enough that the \(LS\)-formalism may break. It is worth noting that many relatively strong lines from heavy elements in the line list of \cite{domoto2022lanthanide} do not satisfy the \(LS\)-coupling rules, such as Ce III \(\lambda \,13,342\,\text{\AA}\) and Th III \(\lambda \,19,947\,\text{\AA}\) lines.

\begin{figure}[ht!]

    \includegraphics[width=0.94\columnwidth]{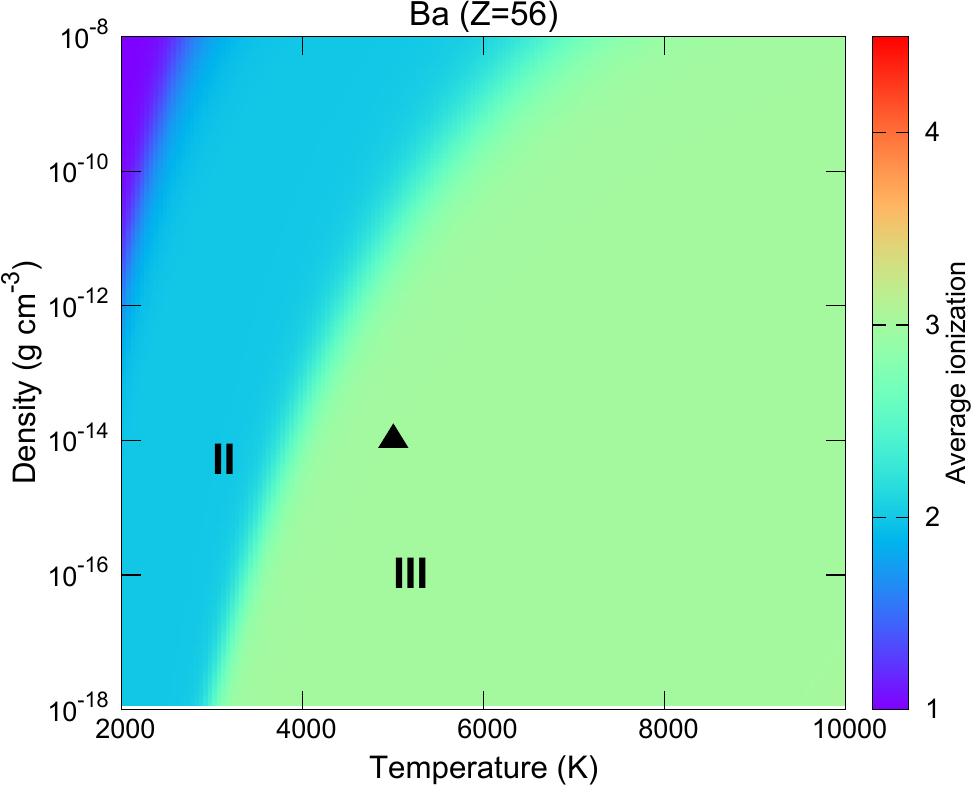} \\
    \includegraphics[width =0.94\columnwidth]{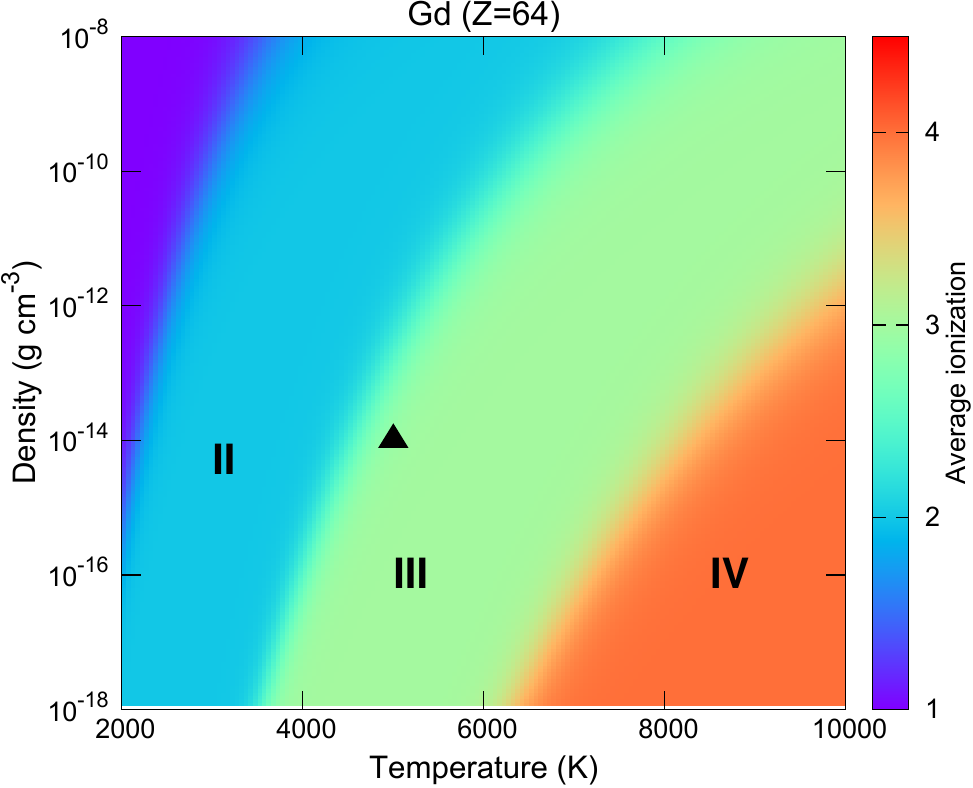} \\
    \includegraphics[width =0.94\columnwidth]{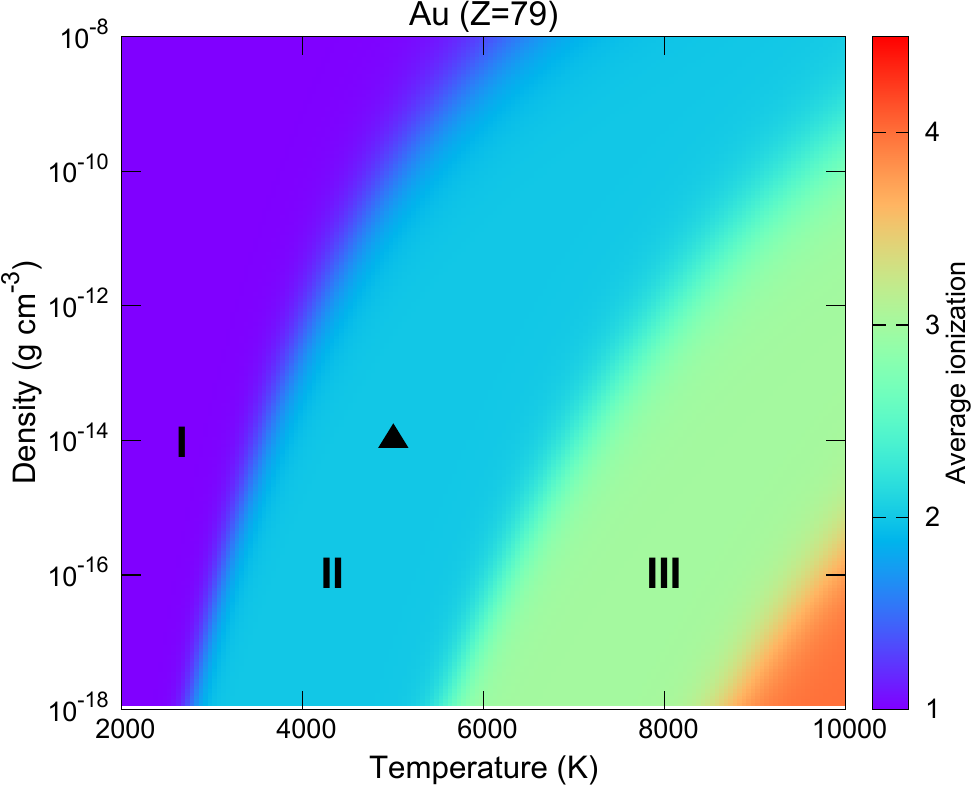}

    \caption{Average ionization as a function of temperature and density for Ba, Gd, and Au from top to bottom respectively. Colors from purple to orange show the neutral (I), singly (II), doubly (III), and triply (IV) ionized states, indicated by 1, 2, 3, and 4, respectively. The triangle symbol highlights typical conditions of the kilonova ejecta a few days after the merger.
    \label{fig:ion}}
\end{figure}

For this study, we only extracted singly and doubly ionized elements from the databases.
Figure \ref{fig:ion} shows the average ionization as a function of temperature and density for three different elements. The ionization is calculated by assuming LTE and solving Saha equation. The assumed abundance pattern is shown in Figure \ref{fig:abun} (see Section \ref{sec:rad}).
We chose Ba (\(Z=56\)), Gd (\(Z=64\)), and Au (\(Z=79\)) in particular because they belong to distinct groups in the periodic table and have different valencies and ionization energies.
These elements are typically found to be singly to doubly ionized for a temperature of \(T=5000\,\)K and matter density of \(\rho=10^{-14}\,\text{g cm}^{-3}\), which are typical conditions of the kilonova atmosphere a few days after the merger. Since the ionization behavior is similar across each group, other elements are also singly or doubly ionized at early time.

Deviations from LTE are considered negligible at early time \citep{kasen2013opacities}, which makes the LTE approximation reasonable. However, if we were to consider non-LTE effects, we would expect a higher ionization fraction due to non-thermal ionizations in the ejecta \citep{hotokezaka2021nebular,pognan2022b,pognan2023nlte}.
This suggests that neutral elements are unlikely to exist at early time. But it also indicates that we could expect a higher fraction of triply ionized elements. However, we did not include these due to the lack of data on triply ionized heavy elements in the NIST ASD and the SCASA database.

Another constraint we imposed for our selection is a threshold of \(2\,\text{eV}\) on the lower energy levels of the selected transitions. The strength of a bound-bound transition in the kilonova ejecta can be calculated using its Sobolev optical depth (e.g. \citealt{barnes2013effect,tanaka2013radioactively,domoto2022lanthanide}). Assuming LTE and a Boltzmann distribution of the population of excited states, the Sobolev optical depth of a certain line can be calculated as
\begin{equation}\label{eq:sobolev}
    \tau_l=\frac{\pi e^2}{m_e c^2}\,n_{i,j}t\,\lambda_l\,f_l\frac{g_l}{Z}e^{-\frac{E_l}{kT}}
\end{equation}
where \(g_l\), \(E_l\), and \(f_l\) are the statistical weight, the energy of the lower level, and the oscillator strength of the bound-bound transition, respectively. \(Z=\sum_ig_i\,e^{-\frac{E_i}{kT}}\) is the atomic partition function and \(n_{i,j}\) is the number density of the \(i\)-th element at the \(j\)-th ionization state.

A transition becomes strong if it has a high oscillator strength (high \(gf\)-value) and a low energy \(E_l\). For example, the strongest line-forming transitions in NIR La III \(\lambda\,14,100\,\text{\AA}\) and Ce III \(\lambda\,15,851\,\text{\AA}\) have
\(\log gf=-0.587\) and \(-0.613\) with a lower energy of transition \(E_l=0.199\,\text{eV}\) and \(0.189\,\text{eV}\) respectively. Considering a typical ejecta temperature of \(T=5000\,\text{K}\), a transition from a level of energy \(E_l=2\,\text{eV}\) is \(e^{-\frac{E_l}{kT}}\approx0.009\) weaker than a transition occurring from the ground level for a given \(gf\)-value. This two orders of magnitude difference in strength is unlikely to be compensated by a high \(gf\)-value since the strongest transitions have an overall \(\log gf \lesssim 0\) with \(e^{-\frac{E_l}{kT}}\sim1\).
Therefore, 2 eV can be considered a good energy threshold for the purpose of our study.

\cite{gillanders2024modelling} carried out a similar analysis using the NIST ASD to search for candidate elements with allowed transitions that coincide with the emission and absorption features in the spectra of AT2017gfo. One key difference between our studies is that \cite{gillanders2024modelling} used looser selection criteria by setting a higher energy threshold of \(4\,\text{eV}\), and including neutral elements in their analysis. In contrast, our stricter criteria significantly narrow the list of candidate elements, focusing on the strongest elements likely to exhibit transitions at early time. Another main difference is that our study includes a comprehensive search of actinide (\(89\leq Z \leq103\)) transitions using the energy levels in the SCASA database.

\begin{figure*}[ht!]

\begin{tabular}{cc}
        \includegraphics[width=0.95\columnwidth]{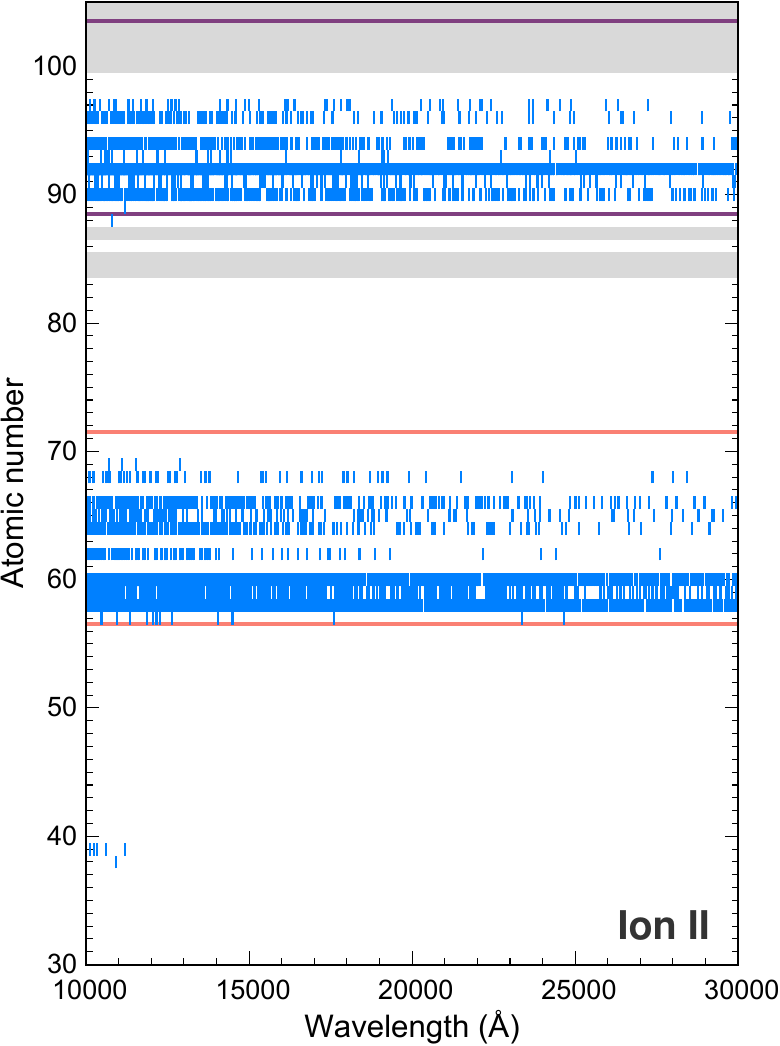} & \includegraphics[width = 0.95\columnwidth]{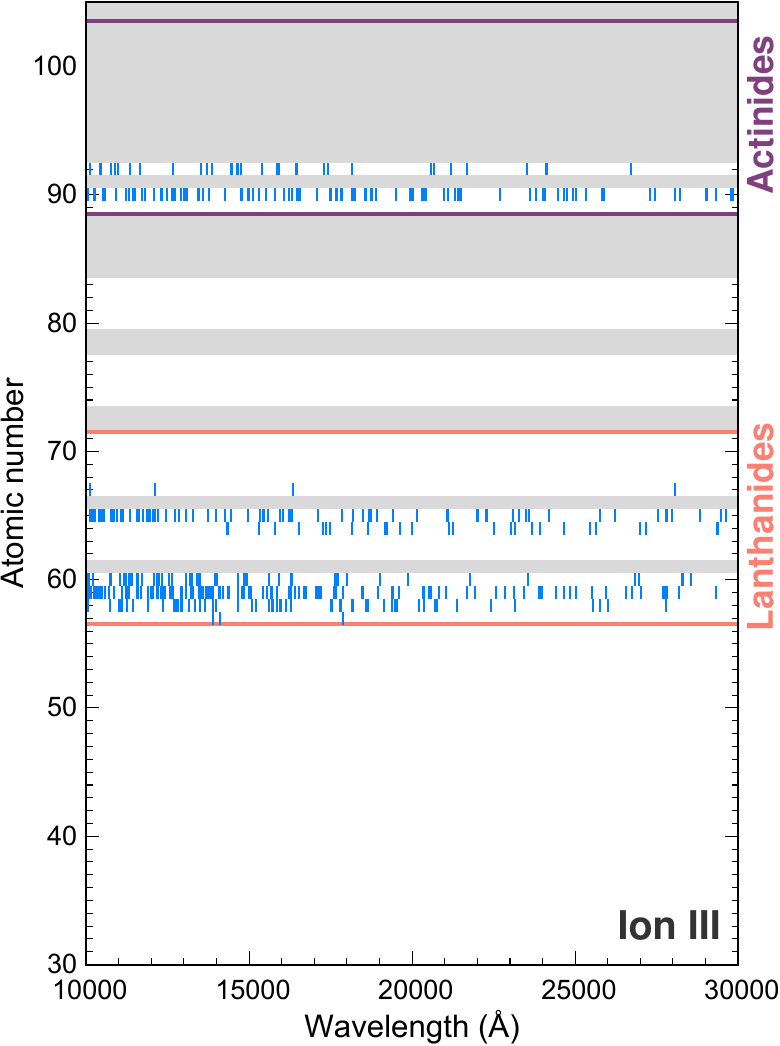}
\end{tabular}
\caption{Left: Wavelengths of allowed NIR transitions occurring below an energy threshold of
2 eV for singly ionized states. 
Elements excluded are shown as gray areas. Lanthanides (\(57\leq Z \leq 71\)) and actinides (\(89\leq Z \leq 103\)) are highlighted in red and purple, respectively. Right: Same as left panel but for doubly ionized states.
\label{fig:cand}}
\end{figure*}

The selection of strong NIR transitions with this method provides an accurate line list for heavy elements. However, its main limitation is that energy levels in both the NIST ASD and the SCASA database are not fully available for all heavy elements.
We have thus decided to exclude elements for which the maximum available energy on the database is below the \(2\,\text{eV}\) threshold. Singly ionized elements excluded are \(Z=\)84, 85, 87. Doubly ionized elements excluded are \(Z=\)61, 66, 72, 73, 78, 79, 84, 85, 86, 87, 88 and all actinides except \(Z=\)89, 90 and 92. All ions with \(Z\geq100\) are excluded due to lack of data (see Figure \ref{fig:cand}).

\subsection{Results}\label{subsec:cand_res}

The wavelengths of the selected NIR transitions for all the elements at singly and doubly ionized states are shown in Figure \ref{fig:cand}. We find that many candidate elements are lanthanides (\(Z=57-71\)), with most light elements (\(Z<57\)) not exhibiting any strong NIR transitions. Sr II (\(Z=38\)) and Y II (\(Z=39\)) are exceptions, with a few transitions at \(\sim10,000\,\text{\AA}\).
Elements between lanthanides and actinides (\(Z=72-88\)) also show no transitions, except for a single line of Ra II (\(Z=88\)). For doubly ionized states, there is a notable lack of data for many elements in this atomic number range, as indicated by gray-shaded areas. Figure \ref{fig:cand} also shows that many singly ionized actinides (\(Z=89-103\)) are important candidate elements. On the other hand, the only doubly ionized actinides for which energy level data are available are Ac, Th, and U (\(Z=\)89, 90, and 92, respectively), out of which Th III and U III show many transitions across the NIR range.

The significance of lanthanides and actinides stems from their atomic structures. The dominant open shell of lanthanides is \(4f\), but their structure also involves the \(5d\) and \(6s\) orbitals. Similarly, actinides' atomic structure includes \(5f\), \(6d\) and \(7s\) as open shells. These complex configurations result in many low-lying energy levels for lanthanides and actinides, leading to strong allowed transitions in the NIR range.

\begin{figure*}[ht!]
        \centering
        \includegraphics[width = \textwidth]{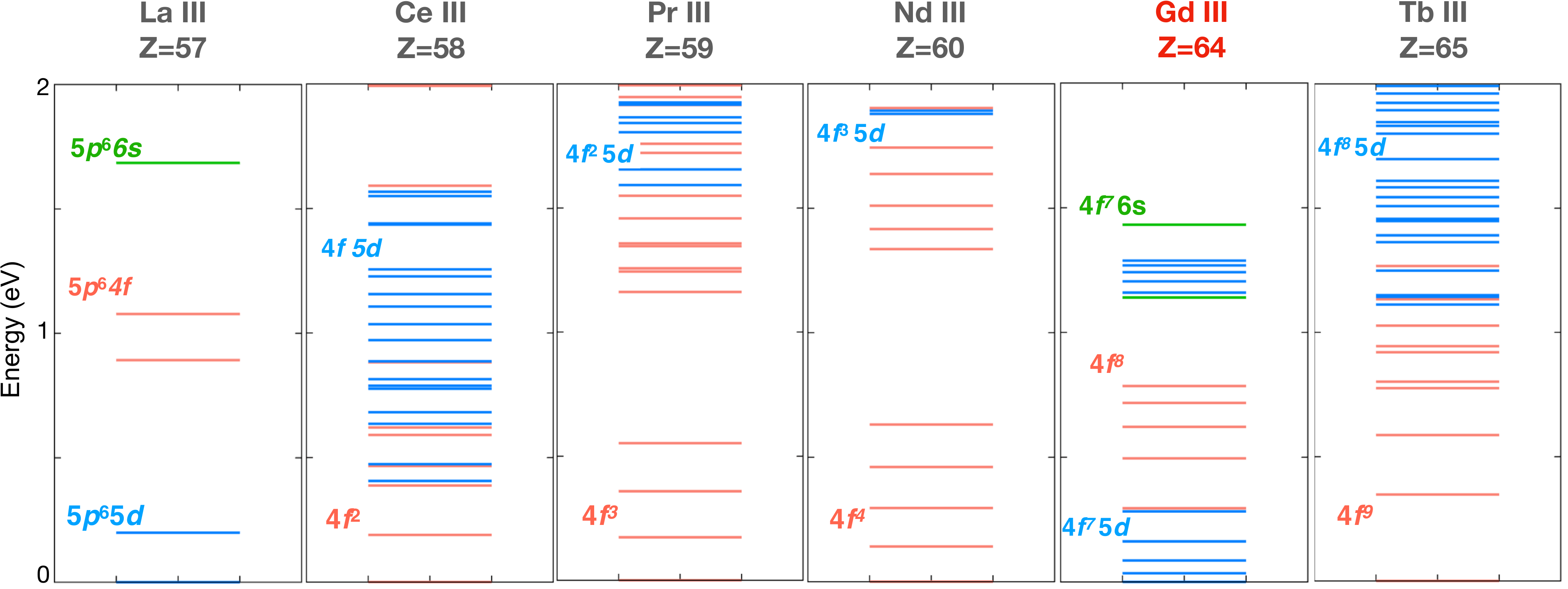}
    \caption{Energy levels below 2 eV for key doubly ionized candidate lanthanides.
    The levels for all elements are sourced from the NIST ASD \citep{nist}, except for Nd III, which is taken from \cite{ding2024spectrum}.
    The red, blue, and green energy levels represent the atomic state where the outer electron is in the \(4f\), \(5d\), and \(6s\) orbitals, respectively. Gd III and La III are the only lanthanides with their ground state outer electron in the \(5d\) orbital, resulting in low-lying energy levels. \label{fig:e_lev}}
    \end{figure*}

Sr II, Y II, and Ra II show a few transitions at shorter wavelengths of the NIR range. These elements, located on the left side of the periodic table, have few valence electrons and low-lying energy levels. However, the energy gaps between these levels are larger than those of lanthanides or actinides, resulting in only a few lines at shorter wavelengths. The importance of Sr II, Y II, and Ra II has been discussed in previous studies \citep{watson2019identification, domoto2022lanthanide, sneppen2023discovery, domoto2024}.

Our list of candidate elements includes La III and Ce III, which is consistent with the results of \cite{domoto2022lanthanide}.
\cite{gillanders2024modelling} used a similar method and identified viable candidate species with strong NIR transitions including Gd III and Ac I.
While the significance of Gd III is further discussed in the next section, Ac I is not included in our results, as neutral atoms are unlikely to exist at early time.

\section{Gd III as a candidate} \label{sec:gd}

After narrowing down the elements with significant transitions in the NIR range in Section \ref{sec:cand}, it is now necessary to investigate which is most likely to exhibit absorption features in the spectrum. In this section, we explain how Gd III is one of the most promising elements to appear in the NIR spectrum of kilonovae.

\subsection{Atomic structure} \label{subsec:gd_atom}

The ground electronic configuration of Gd III is \(\left[\text{Xe}\right]4f^75d\). It is the only doubly ionized lanthanide, aside from La III, with electrons in the \(5d\) orbital in its ground state.
Unpaired (parallel spin) electron configurations are energetically favorable (lower) because the parallel spin electrons cannot be situated at identical spatial positions due to Pauli exclusion principle, having smaller electrostatic repulsion. The \(4f\) orbital in Gd III contains 7 unpaired electrons, which is the largest number of unpaired electrons in the \(4f\) orbital. This lowers the energy of the atom as a whole. Adding an 8th electron to the \(4f\) orbital decreases the number of unpaired electrons and results in a higher energy state than adding the electron to the \(5d\) orbital.
Consequently, Gd III has lower energy levels compared to other doubly ionized candidate elements, as shown in Figure \ref{fig:e_lev}. Different colors in Figure \ref{fig:e_lev} represent different electronic configurations. Transitions generally occur between two different atomic states, and stronger absorption features tend to originate from lower, more populous energy levels (see Equation \ref{eq:sobolev}). Therefore, the unique atomic structure of Gd III makes it a strong candidate for further investigation.

\cite{domoto2022lanthanide} did not include Gd III in their analysis because theoretical calculations often overestimate energy levels. Moreover,  the theoretically calculated transition probabilities of Gd III NIR lines tend to be low and negligible, as will be further discussed in Section \ref{subsec:gd_line}.
It is generally expected that the elements most likely to show strong lines belong to the left side of the periodic table due to their small number of valence electrons and relatively low-lying energy levels. While Gd is located around the middle of the periodic table, its doubly ionized electron configuration involves the \(5d\) orbital as a valence shell with only one electron. This makes Gd III an exception to the general trend.

\begin{figure}[t!]
\includegraphics[width=\columnwidth]{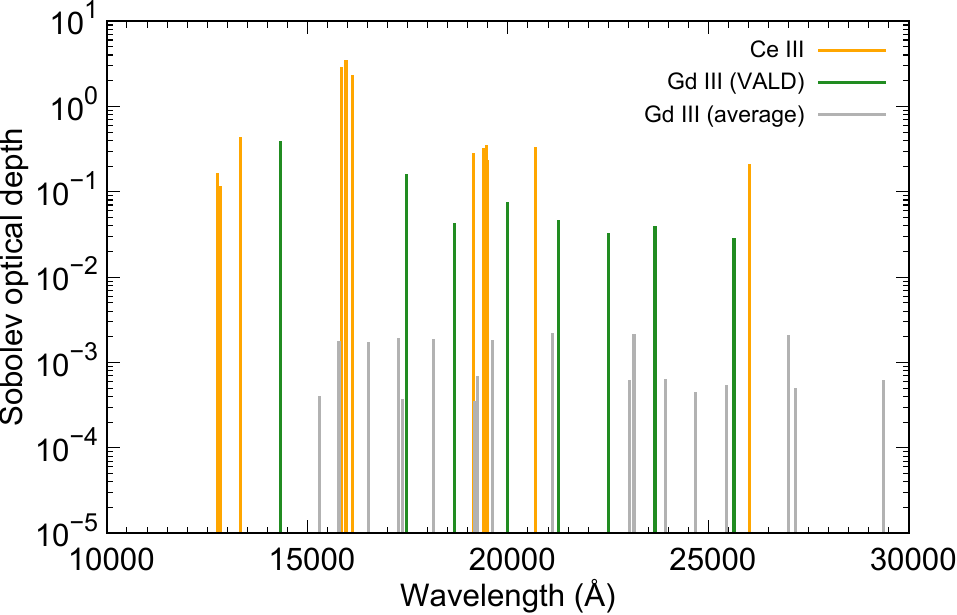}
\caption{Sobolev optical depth of Ce III and Gd III lines assuming \(T=5000\,\text{K}\) and \(\rho=10^{-14}\,\text{g cm}^{-3}\). Gd III lines are plotted using the atomic data shown in Table \ref{tab:gd}, where lines with the empirical VALD \(gf\)-values and the theoretically calculated average \(gf\)-value are shown in green and gray, respectively. Ce III lines are shown in orange, and the element's atomic data are taken from the calibrated theoretical data in \citet{domoto2022lanthanide}. Only lines with a Sobolev optical depth higher than 0.1 are plotted for Ce III. \label{fig:tau_cegd}}
\end{figure}

\begin{deluxetable*}{cccccc}
\tablenum{1}
\tablecaption{NIR transition lines of Gd III\label{tab:gd}}
\tablewidth{0pt}
\tablehead{
\colhead{\(\lambda(\text{\AA})^a\)} & \colhead{Lower level} & \colhead{\(E_{\rm{lower}}(\text{eV})^b\)} & \colhead{Upper level} & \colhead{\(E_{\rm{upper}}(\text{eV})^c\)} & \colhead{\(\log gf^d\)}
}
\startdata
14336.78 & \(4f^8\ ^7\text{F}_6\) & 0.295 & \(4f^7(^8\text{S}^{\circ})5d\ ^7\text{D}^{\circ}_5\) & 1.160 & -1.521 \\
15303.86 & \(4f^8\ ^7\text{F}_4\) & 0.622 & \(4f^7(^8\text{S}^{\circ})6s\ ^7\text{S}^{\circ}_3\)  & 1.432 & \(-\) \\
15787.39 & \(4f^7(^8\text{S}^{\circ})5d\ ^9\text{D}^{\circ}_2\) & 0 & \(4f^8\ ^7\text{F}_2\)  & 0.785 & \(-\) \\
16515.68 & \(4f^7(^8\text{S}^{\circ})5d\ ^9\text{D}^{\circ}_3\) & 0.035 & \(4f^8\ ^7\text{F}_2\)  & 0.785 &  \(-\) \\
17271.24 & \(4f^7(^8\text{S}^{\circ})5d\ ^9\text{D}^{\circ}_2\) & 0 & \(4f^8\ ^7\text{F}_3\)  & 0.718 &  \(-\) \\
17362.52 & \(4f^8\ ^7\text{F}_3\) & 0.718 & \(4f^7(^8\text{S}^{\circ})6s\ ^7\text{S}^{\circ}_3\)  & 1.432 &  \(-\) \\
17479.48 & \(4f^8\ ^7\text{F}_5\) & 0.496 & \(4f^7(^8\text{S}^{\circ})5d\ ^7\text{D}^{\circ}_4\)  & 1.205 & -1.791 \\
18146.67 & \(4f^7(^8\text{S}^{\circ})5d\ ^9\text{D}^{\circ}_3\) & 0.035 & \(4f^8\ ^7\text{F}_3\)  & 0.718 & \(-\) \\
18658.13 & \(4f^8\ ^7\text{F}_5\) & 0.496 & \(4f^7(^8\text{S}^{\circ})5d\ ^7\text{D}^{\circ}_5\)  & 1.160 & -2.399 \\
19174.23 & \(4f^8\ ^7\text{F}_2\) & 0.785 & \(4f^7(^8\text{S}^{\circ})6s\ ^7\text{S}^{\circ}_3\)  & 1.432 & \(-\) \\
19236.94 & \(4f^8\ ^7\text{F}_5\) & 0.496 & \(4f^7(^8\text{S}^{\circ})6s\ ^9\text{S}^{\circ}_4\)  & 1.140 & \(-\) \\
19624.77 & \(4f^7(^8\text{S}^{\circ})5d\ ^9\text{D}^{\circ}_4\) & 0.086 & \(4f^8\ ^7\text{F}_3\)  & 0.718 &  \(-\) \\
20001.80 & \(4f^8\ ^7\text{F}_4\) & 0.622 & \(4f^7(^8\text{S}^{\circ})5d\ ^7\text{D}^{\circ}_3\)  & 1.242 & -2.057 \\
21115.37 & \(4f^7(^8\text{S}^{\circ})5d\ ^9\text{D}^{\circ}_3\) & 0.035 & \(4f^8\ ^7\text{F}_4\)  & 0.622 & \(-\) \\
21265.27 & \(4f^8\ ^7\text{F}_4\) & 0.622 & \(4f^7(^8\text{S}^{\circ})5d\ ^7\text{D}^{\circ}_4\)  & 1.205 & -2.293 \\
22499.58 & \(4f^8\ ^7\text{F}_3\) & 0.718 & \(4f^7(^8\text{S}^{\circ})5d\ ^7\text{D}^{\circ}_2\)  & 1.269 & -2.370 \\
23035.63 & \(4f^8\ ^7\text{F}_4\) & 0.622 & \(4f^7(^8\text{S}^{\circ})5d\ ^7\text{D}^{\circ}_5\)  & 1.160 & \(-\) \\
23143.69 & \(4f^7(^8\text{S}^{\circ})5d\ ^9\text{D}^{\circ}_4\) & 0.086 & \(4f^8\ ^7\text{F}_4\)  & 0.622 & \(-\) \\
23669.87 & \(4f^8\ ^7\text{F}_3\) & 0.718 & \(4f^7(^8\text{S}^{\circ})5d\ ^7\text{D}^{\circ}_3\)  & 1.242 & -2.318 \\
23924.37 & \(4f^8\ ^7\text{F}_4\) & 0.622 & \(4f^7(^8\text{S}^{\circ})6s\ ^9\text{S}^{\circ}_4\)  & 1.140 & \(-\) \\
24672.24 & \(4f^8\ ^7\text{F}_2\) & 0.785 & \(4f^7(^8\text{S}^{\circ})5d\ ^7\text{D}^{\circ}_1\)  & 1.288 &  \(-\) \\
25459.97 & \(4f^8\ ^7\text{F}_3\) & 0.718 & \(4f^7(^8\text{S}^{\circ})5d\ ^7\text{D}^{\circ}_4\)  & 1.205 & \(-\) \\
25638.87 & \(4f^8\ ^7\text{F}_2\) & 0.785 & \(4f^7(^8\text{S}^{\circ})5d\ ^7\text{D}^{\circ}_2\)  & 1.269 & -2.421 \\
26989.99 & \(4f^7(^8\text{S}^{\circ})5d\ ^9\text{D}^{\circ}_5\) & 0.162 & \(4f^8\ ^7\text{F}_4\)  & 0.622 & \(-\) \\
27169.62 & \(4f^8\ ^7\text{F}_2\) & 0.785 & \(4f^7(^8\text{S}^{\circ})5d\ ^7\text{D}^{\circ}_3\)  & 1.242 & \(-\) \\
29367.96 & \(4f^8\ ^7\text{F}_3\) & 0.718 & \(4f^7(^8\text{S}^{\circ})6s\ ^9\text{S}^{\circ}_4\)  & 1.140 & \(-\)  \\
\enddata
\tablecomments{\(^a\) Vacuum transition wavelength \\ \(^b\) Lower energy level \\
\(^c\) Upper energy level \\ \(^d\) logarithm of \(gf\)-value from the VALD database
\citep{piskunov1995vald, kupka1999vald, ryabchikova2015major}}
\end{deluxetable*}

\subsection{Strength of Absorption Features}\label{subsec:gd_line}

To confirm the significance of Gd III lines for absorption features, it is important to evaluate their Sobolev optical depth (see Equation \ref{eq:sobolev}). To do this, we require not only the lower energy levels but also accurate \(gf\)-values.
Theoretical calculations with HULLAC \citep{tanaka2020systematic} tend to overestimate the energy levels of heavy elements, and a calibration is necessary to find the true \(gf\)-value of each transition.
While \cite{domoto2022lanthanide} calibrated the energy levels of several key elements, the calibration for Gd III is challenging due to its complex structure. Overall, we find that the theoretical calculations for Gd III give an average \(\log gf=-5.7\).
This is significantly lower than the average \(\log gf=-1.9\) for Ce III \citep{tanaka2020systematic}.
Gd III has a large complexity due to its high number of energy levels \citep{kasen2013opacities}. This makes the transition probability of each line smaller, causing a low average \(gf\)-value.
Nevertheless, several empirical calculations were performed for Gd III, and the \(gf\)-values of some of its transitions are listed in the VALD database \citep{piskunov1995vald, kupka1999vald, ryabchikova2015major}.
These empirical values were calculated from the emission spectra of Gd III via Cowan code \citep{cowan1981theory}, in the same way that previously recorded by \cite{ding2024spectrum} for Nd III (\citealt{ISAN}, private communication). The VALD \(gf\)-values of Gd III are significantly higher than the theoretical average calculated by \cite{tanaka2020systematic}, highlighting the underestimation of the theoretical calculations to the true \(gf\)-values. Table \ref{tab:gd} lists all allowed transitions of Gd III in the NIR wavelength range, selected in Section \ref{sec:cand}, along with the \(gf\)-values of lines available in the VALD. As explained in Section \ref{sec:cand}, the energies and configurations are taken from the NIST ASD.

Using the values in Table \ref{tab:gd}, we calculated the Sobolev optical depth for Gd III lines. The ion number density \(n_{i,j}\) was obtained by solving Saha equation and assuming the abundance pattern as in Figure \ref{fig:abun}.
A typical temperature of \(T=5000\,\text{K}\) and density \(\rho = 10^{-14}\,\text{g cm}^{-3}\) were assumed. Figure \ref{fig:tau_cegd} compares the Sobolov optical depths of Gd III lines and those of the strongest Ce III lines. The theoretical average \(gf\)-value with HULLAC code was used for lines not available in the VALD, as individual theoretically calculated \(gf\)-values are not available.
For Ce III, we used the calibrated theoretical data in Table 4 of \citet{domoto2022lanthanide}, and included only lines with Sobolev optical depth \(\tau\geq0.1\).
The strongest Gd III line, at \(\lambda = 14,336\,\text{\AA}\), shows a comparable Sobolev optical depth to Ce III lines. This suggests that the impact of Gd III (particularly the \(\lambda = 14,336\,\text{\AA}\) line) on the kilonova spectrum is worth investigating.

\subsection{Emergence in The Spectrum of a Chemically Peculiar Star}\label{subsec:gd_cp}

While we have shown that Gd III has low-lying energy levels and relatively high \(gf\)-values, it is important to investigate whether its lines can produce spectral features in kilonovae. The spectra of chemically peculiar stars are one way to explore this. These stars have abnormal elemental abundance patterns, sometimes with extremely enhanced abundances of elements heavier than Fe. This makes their spectra a good astrophysical laboratory to study kilonovae \citep{tanaka2023cerium}. The spectrum of a late B-type chemically peculiar star HR 465 was obtained by \cite{tanaka2023cerium} using the Subaru Telescope IRD instrument \citep{tamura2012infrared,kotani2018infrared}.
They showed that the star exhibits abundance patterns and ionization degrees of lanthanides similar to those in the kilonova ejecta.

A detailed comparison between the absorption lines in HR 465 and the transitions selected in Section \ref{sec:cand} reveals that two spectral features correspond to Gd III \(14,336\,\text{\AA}\) and \(17,479\,\text{\AA}\) lines. Figure \ref{fig:tau_cegd} shows that these two transitions have the highest Sobolev optical depth, meaning that they are the most likely to appear when Gd III is present.
This further suggests that the \(14,336\,\text{\AA}\) and \(17,479\,\text{\AA}\) lines are important in kilonova spectra.

\begin{figure*}[ht!]

\begin{tabular}{cc}
    \includegraphics[width=0.95\columnwidth]{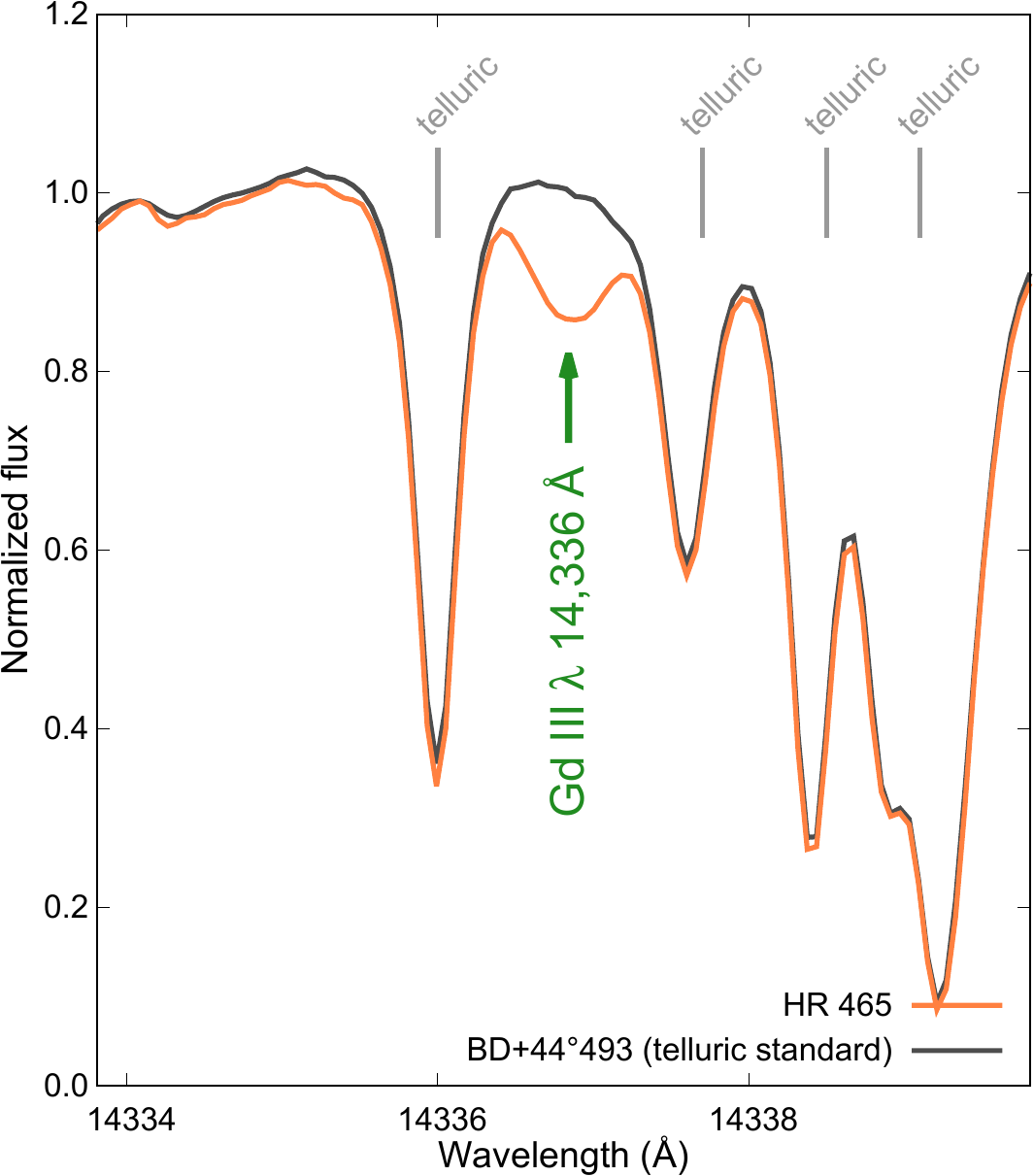} & \includegraphics[width = 0.95\columnwidth]{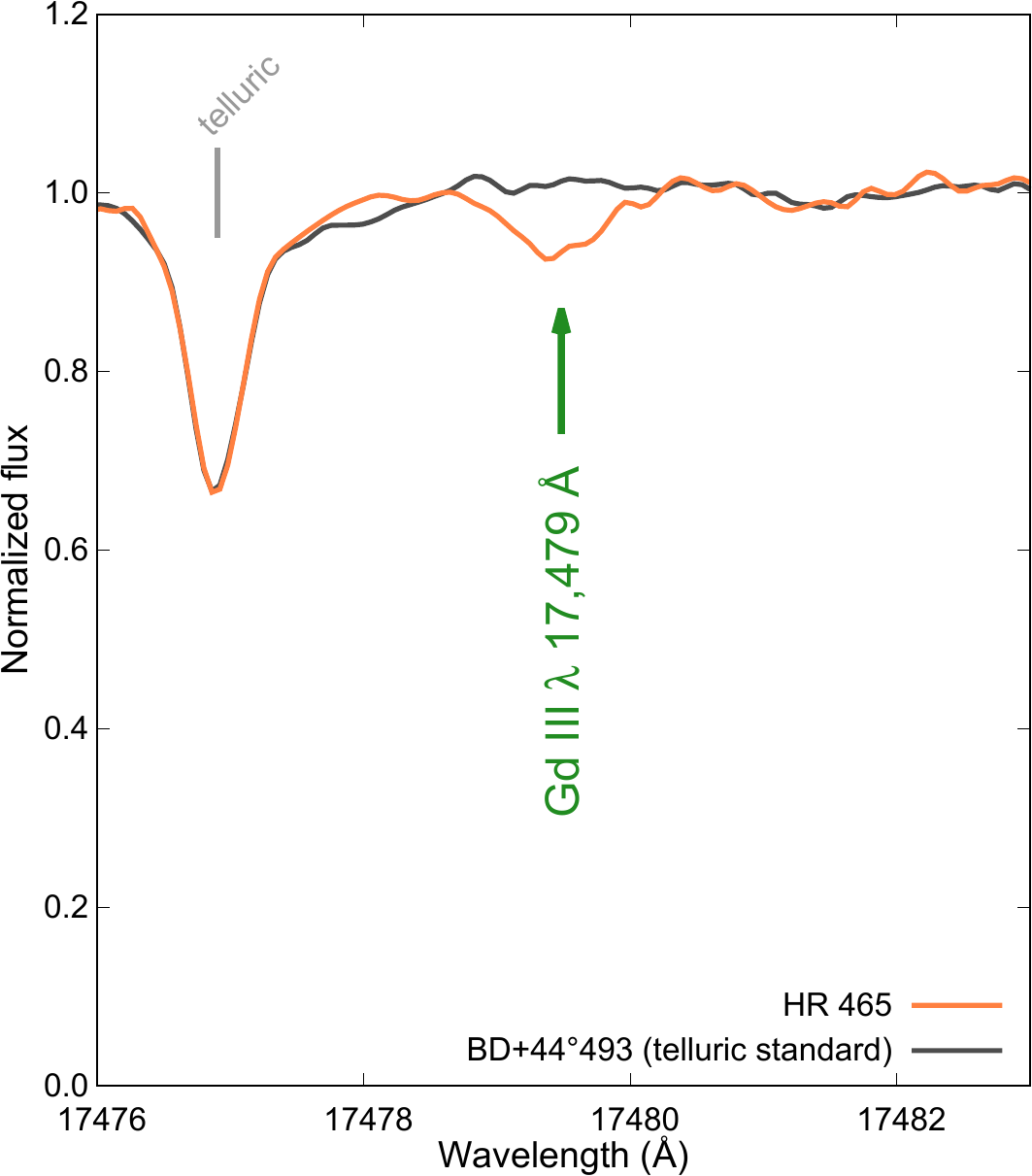} \\
    \includegraphics[width=0.95\columnwidth]{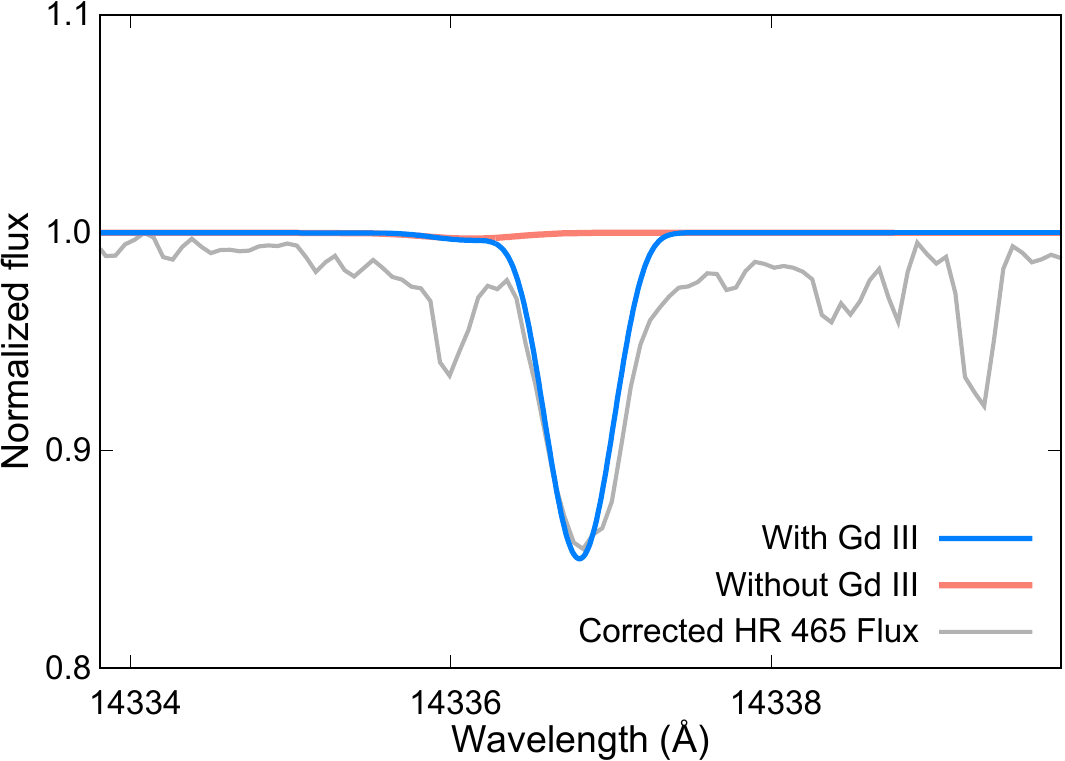} & \includegraphics[width = 0.95\columnwidth]{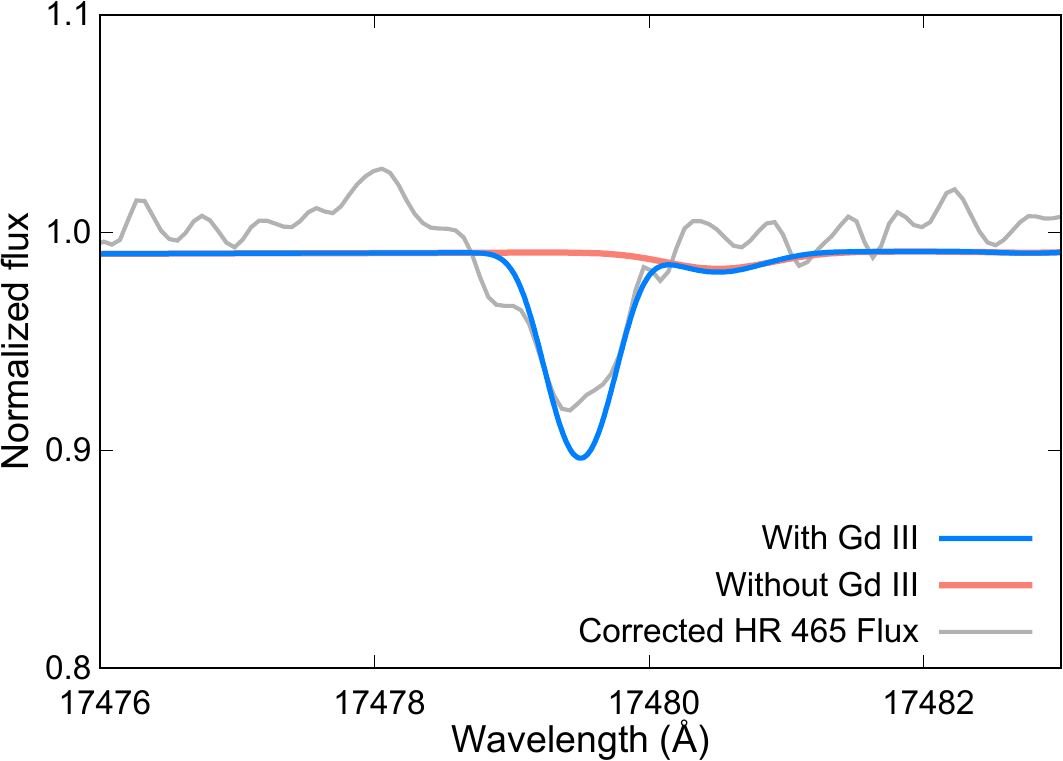}
\end{tabular}
\caption{Top: Uncorrected HR 465 flux and the telluric standard star BD\(+44^\circ 493\) flux around the \(14,336\,\text{\AA}\) line (left) and the \(17,479\,\text{\AA}\) line (right). The flux of BD\(+44^\circ 493\) is corrected for the difference in airmass between the two stars. The telluric absorption features due to the Earth's atmosphere are highlighted by gray lines.
Bottom: Comparison of the normalized corrected HR 465 flux, synthetic spectrum without Gd III, and the best-fit synthetic spectrum with abundance \(\left[\rm{Gd}/\rm{H}\right] = 4.0\) around the \(14,336\,\text{\AA}\) line (left) and the \(17,479\,\text{\AA}\) line (right). The features around the two Gd III absorption lines are due to the residual of telluric correction
\label{fig:pstar}}
\end{figure*}

To further investigate the behavior of Gd III lines, we calculated a model spectrum for HR 465 using Turbospectrum \citep{plez2012turbospectrum} implemented through iSpec \citep{blanco2019modern, blanco2014determining}. We assumed an effective temperature of \(T=11,000\,\text{K}\) and surface gravity of \(\log g=4.0\) based on the derived stellar atmosphere parameters of HR 465 in \citet{nielsen2020advanced}. For simplicity, the microturbulent velocity was set at \(\zeta = 2.0\,\text{km s}^{-1}\), and the macroturbulent velocity was assumed to be \(v=0\,\text{km s}^{-1}\). We used ATLAS9 as the atmospheric model \citep{castelli2003round}. For the atomic data, we used the VALD \citep{piskunov1995vald, kupka1999vald, ryabchikova2015major} line list as a baseline, to which we added Gd III lines from Table \ref{tab:gd}.

We based the Gd abundance in the model on the measured values for HR 465 reported by \cite{nielsen2020advanced}. The abundance pattern of HR 465 is known to vary with a period of 21.5 years, during which lanthanides reach a maximum at phase \(\phi=0\) and a minimum at phase \(\phi=0.5\) \citep{pyper2017photometric}. \citet{nielsen2020advanced} have measured the elemental abundances in HR 465 at phases \(\phi =\) 0.45, 0.68, and 0.85. Gd's abundance, inferred from both Gd II and Gd III lines, was found to be \([\rm{Gd}^{2+}/\rm{H}]=3.2\) at \(\phi=0.45\) and \([\rm{Gd}^+/\rm{H}]=\) 3.3, 3.1, and 5.1 at \(\phi = \) 0.45, 0.68, and 0.85, respectively. Although it is not straightforward to infer \([\rm{Gd}/\rm{H}]\) for the spectrum used in this study, as it was observed at \(\phi=0.77\) \citep{tanaka2023cerium}, the Gd abundance is likely to range from 3.2 to 5.1. 

The bottom panels of Figure \ref{fig:pstar} show the comparison between the normalized HR 465 flux and the synthetic spectra with and without Gd III lines, where \([\rm{Gd}/\rm{H}]=4.0\) was used. Atmospheric absorption was corrected using BD\(+44^\circ 493\) as a telluric standard star due to its low metallicity and very weak absorption lines \citep{aoki2022silicon}. The top panels of Figure \ref{fig:pstar} show the spectrum of BD\(+44^\circ 493\) with the non-corrected flux of HR 465.
The two spectra were taken back-to-back, and the difference in airmass between the two stars was accounted for.
The \(17,479\,\text{\AA}\) feature is not affected by the atmosphere. This line, previously unidentified by \cite{tanaka2023cerium}, is shown in the bottom right panel of Figure \ref{fig:pstar},  where the absorption line nicely matches the synthetic spectrum calculated with Gd III lines. On the other hand, the \(14,336\,\text{\AA}\) line appears in the telluric region and is not discussed by \cite{tanaka2023cerium}. The top left panel of Figure \ref{fig:pstar} shows that the line coincides with a region where the flux of BD\(+44^\circ 493\) has no absorption, and it is very unlikely that the feature is caused by the Earth's atmosphere. Since both lines can be modeled using the VALD \(gf\)-values and assumed abundance of \([\rm{Gd}/\rm{H}]=4.0\), we can conclude that the two lines are indeed Gd III lines. Besides Ce III and Sr II \citep{tanaka2023cerium}, Gd III is the only candidate species appearing in the chemically peculiar star's NIR spectrum, suggesting that it may affect the kilonova spectra.

This result further confirms the reliability of the empirically calculated \(gf\)-values available in the VALD, as these values provide a good fit to the observed spectral features. Therefore, the HULLAC calculations conducted by \cite{tanaka2020systematic} underestimate the oscillator strengths of Gd III; the average \(gf\)-value is several orders of magnitude lower than the true \(gf\)-values, at least for the transitions with available oscillator strengths in Table \ref{tab:gd} (see Section \ref{subsec:gd_line}).

\section{Radiative Transfer Simulations}\label{sec:rad}

\subsection{Setup}\label{subsec:rad_set}

\begin{figure}[t!]
    \includegraphics[width=\columnwidth]{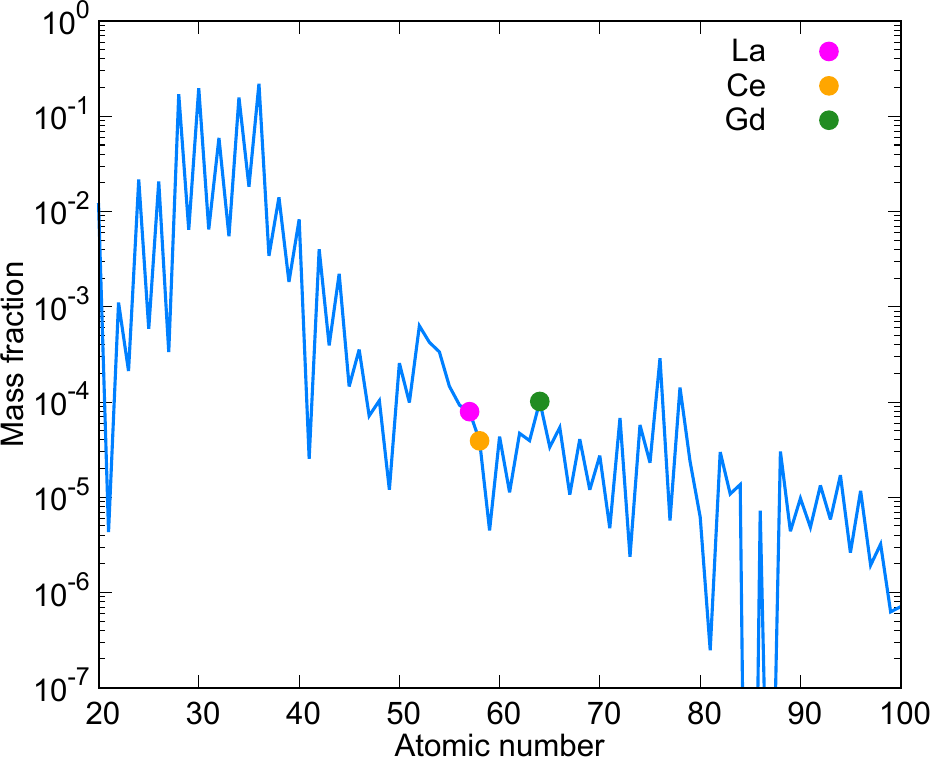}
    \caption{Abundances at 1.5 days after the merger of the Light model (see text) as a function of atomic number. The abundances of La III, Ce III, and Gd III are highlighted as pink, orange, and green circles, respectively.}
    \label{fig:abun}
\end{figure}

To investigate the effect of Gd III lines on the kilonova spectra, we performed radiative transfer simulations using a Monte Carlo radiative transfer code \citep{tanaka2013radiative, kawaguchi2018radiative}. The bound-bound transitions, which predominantly contribute to the opacities in the optical and NIR wavelengths, are expressed using the expansion opacity formalism introduced by \citet{karp1977opacity}, and the formula from \citet{eastman1993spectrum} is used:
\begin{equation}
    \kappa=\frac{1}{ct\rho}\sum_l\frac{\lambda_l}{\Delta \lambda}(1-e^{-\tau_l})
\end{equation}
Here, \(\lambda_l\) and \(\tau_l\) are the wavelength and the Sobolev optical depth of a transition \(l\), respectively. The summation is taken over all transitions in a wavelength bin \(\Delta \lambda\).

We calculated the synthetic spectra using the same setup as \citet{domoto2022lanthanide}.
We assumed a one-dimensional ejecta structure and a homogeneous abundance distribution, with a single power-law density structure (\(\rho\propto r^{-3}\)) for the velocity range \(v=0.05-0.3\,c\). The total ejecta mass is put as \(M_{\rm{ej}}=0.03\,M_{\odot}\), which best explains the observed luminosity of AT2017gfo (e.g., \citealt{tanaka2017kilonova}). \cite{domoto2022lanthanide} adopted a hybrid line list for the atomic data, which includes both complete non-accurate theoretical calculations \citep{tanaka2020systematic}, and calibrated accurate transitions for important elements like La III and Ce III.
The opacity for non-accurate transitions is calculated using a wavelength grid of \(\Delta \lambda = 200\,\text{\AA}\) to avoid significant effects on the spectral features, while for accurate transitions, a finer wavelength grid of \(\Delta \lambda = 10\,\text{\AA}\) is used. We constructed Gd III transition data using the NIST ASD experimentally-constructed energy levels (Section \ref{sec:cand}), which ensures that the transition wavelengths and energy levels are accurate. Therefore, the atomic data of Gd III (Table \ref{tab:gd}) were added as accurate data to the hybrid line list.

The assumed abundance distribution is shown in Figure \ref{fig:abun}. We used the Light abundance model introduced in \cite{domoto2021signatures}, based on the multi-component free-expansion model by \cite{wanajo2018physical}. This abundance model provides a good match with the observed spectrum of AT2017gfo \citep{domoto2022lanthanide}. Figure \ref{fig:abun} shows that lanthanides, particularly La, Ce, and Gd\(-\)elements most important to our study\(-\)have similar mass fractions. This suggests that the relative effect of these elements on the spectrum comes mainly from the strength of their lines, rather than their abundance.

\begin{figure}[t!]
    \includegraphics[width=\columnwidth]{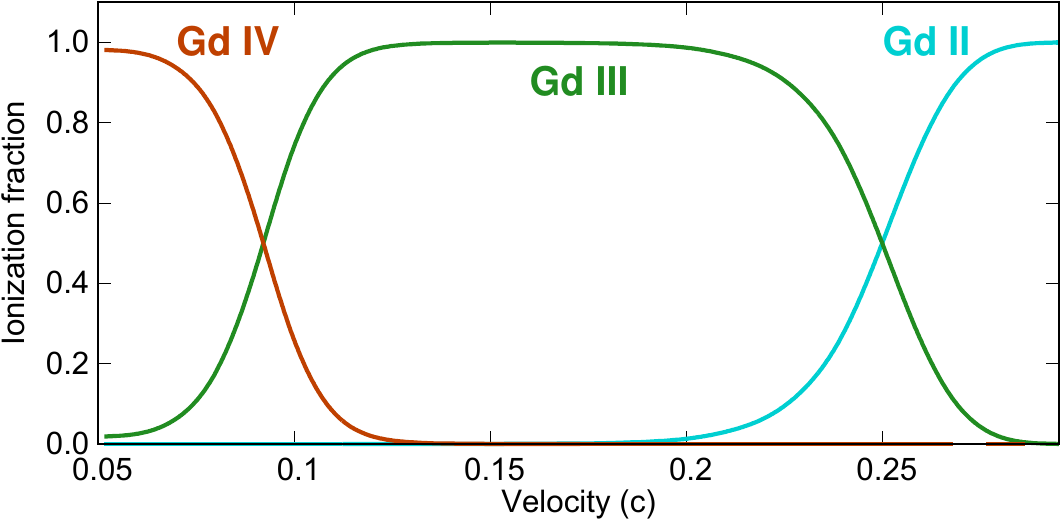}
    \caption{Variation of Gd ionization fraction with the ejecta velocity at 1.5 days after the merger.}
    \label{fig:ion_frac}
\end{figure}

Using the abundance pattern from Figure \ref{fig:abun}, the ionization is calculated by assuming LTE and solving Saha equations. Elements at early time of kilonovae are mainly either singly or doubly ionized, which agrees with our assumptions in Section \ref{subsec:cand_method}. An example is shown in Figure \ref{fig:ion_frac}, where the evolution of Gd ionization fraction throughout the ejecta is plotted for \(t=1.5\) days after the merger. The element is mainly doubly ionized at the NIR line-forming region, found to be at \(v\approx 0.16\,c\) \citep{domoto2022lanthanide}.

\subsection{Results}\label{subsec:rad_res}

\begin{figure*}[ht!]
    \centering
    \includegraphics[width = 0.863\textwidth]{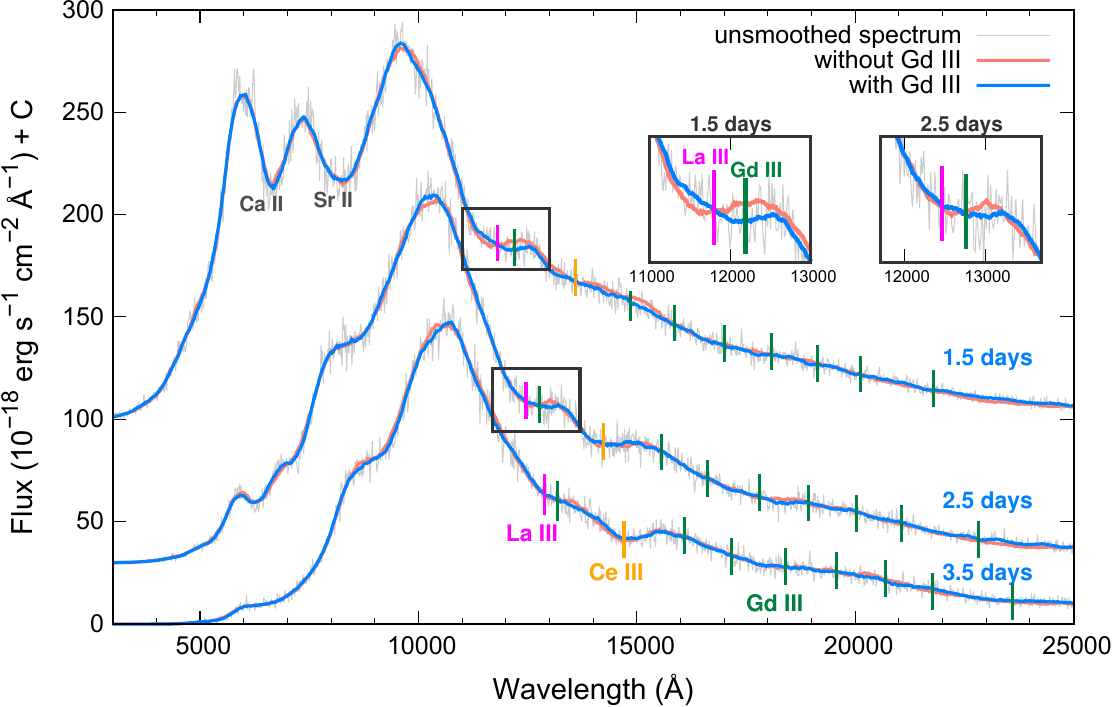}\\
    \includegraphics[width = 0.863\textwidth]{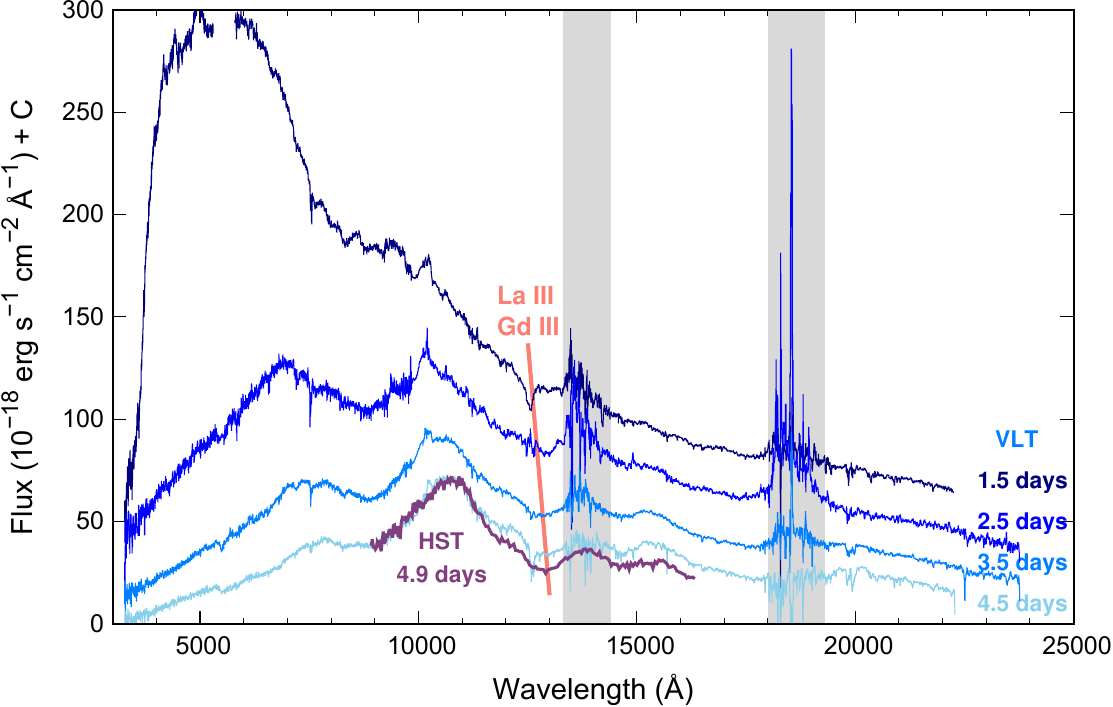}
    \caption{Top: Synthetic spectra of kilonova at 1.5, 2.5, and 3.5 days after the merger from top to bottom, calculated with and without Gd III lines. The spectra are vertically shifted by a constant \(C\) for better visualization with \(C=100\) and 30 for \(t=1.5\) and 2.5 days, respectively. The green ticks show the blueshifted wavelengths of the strongest lines of Gd III (\(v=0.15\,c\), \(0.11\,c\), and \(0.08\,c\) for \(t=1.5\), 2.5, and 3.5 days, respectively). The pink and orange ticks indicate the wavelengths of the feature-forming lines of La III and Ce III respectively. The close-up plots highlight the effect of Gd III \(\lambda\,14,336\,\text{\AA}\) at 1.5 and 2.5 days.
    Bottom: Spectra of AT2017gfo taken with the Very Large Telescope (blue lines, \citealt{pian2017spectroscopic,smartt2017kilonova}) and the Hubble Space Telescope (purple line, \citealt{tanvir2017emergence}). Gray shaded areas show the regions of strong atmospheric absorption. The spectra are shifted with \(C=\)50, 30, and 10 for \(t=\) 1.5, 2.5, and 3.5 days, respectively.}\label{fig:rad_1}
\end{figure*}

The top panel of figure \ref{fig:rad_1} shows the synthetic spectra at 1.5, 2.5, and 3.5 days after the merger from top to bottom, both with and without Gd III lines. The spectra were smoothed using the Savitzky-Golay filter, with a window length of \(\Delta\lambda=25\,\text{\AA}\), and first-order polynomials were applied to fit the spectra within each wavelength window. The two absorption features at \(6,500\,\text{\AA}\) and \(8,000\,\text{\AA}\) are attributed to Ca II and Sr II, respectively, and were previously investigated by \cite{domoto2021signatures}. The green ticks mark the blueshifted wavelengths of the Gd III lines from the VALD transition data, as these lines have the highest \(gf\)-values and are the most likely to impact the spectra (see Table \ref{tab:gd}). The blueshifts were calculated based on the Ce III feature at \(\lambda\sim 14,000\,\text{\AA}\), which gave values of \(0.15\,c\), \(0.11\,c\), and \(0.08\,c\) at 1.5, 2.5, and 3.5 days after the merger, respectively.

The close-up plots in the top panel of Figure \ref{fig:rad_1} highlight the wavelength ranges where Gd III has the strongest effect at 1.5 and 2.5 days after the merger. The shift observed between synthetic spectra with and without Gd III corresponds to the line with the highest \(gf\)-value and the largest Sobolev optical depth (see Figure \ref{fig:tau_cegd}). It also aligns with the observed feature at \(\lambda\sim 12,000\,\text{\AA}\) that was previously attributed to La III \citep{domoto2022lanthanide}. After including Gd III lines, the absorption feature becomes more prominent and its center shifts by about \(500\,\text{\AA}\) at 1.5 days. However, the effect of Gd III decreases with time and becomes negligible at around 3.5 days. At this point, La III remains the main element explaining the absorption feature.

\section{Discussions}\label{sec:disc}

By calculating the synthetic spectra using Gd III atomic data, we found that the \(14,336\,\text{\AA}\) line could give a notable contribution to the kilonova absorption feature at \(12,000\,\text{\AA}\), particularly at early time. This feature was previously attributed to La III \citep{domoto2022lanthanide}. We found that the presence of Gd III causes the absorption to become broader and its center to shift by approximately \(500\,\text{\AA}\) at 1.5 days after the merger.

The ionization energy of Gd II is 12.1 eV, while that of La II is 11.2 eV \citep{nist}. Therefore, as the kilonova ejecta expands and cools, we expect Gd to become singly ionized before La.
This would cause the absorption to change shape and become less prominent, as can be observed in the top panel of Figure \ref{fig:rad_1}. By 3.5 days after the merger, Gd III's contribution diminishes as it becomes singly ionized, leaving La III as the primary element causing the feature.
High-cadence NIR observations of future kilonovae would allow us to follow the evolution of this feature and confirm the effects of Gd III lines.
Simulations of neutron star mergers predict that the mass fractions across lanthanides are relatively similar (e.g, \citealt{fujibayashi2023}). Since Gd has a higher atomic number compared to La and Ce, constraining its relative abundance from the shifted observed feature could provide further tests of \(r\)-process abundance patterns in neutron star mergers.

As the ejecta expands and its density decreases, deviations from LTE become more prominent. This causes the temperature structures and ionization fractions to be different from those used in the synthetic spectra in Section \ref{sec:rad}. While the LTE approximation is considered reasonable at around 1.5 days after the merger \citep{kasen2013opacities}, non-thermal processes might keep Gd in a doubly ionized state for longer than 3.5 days. An estimation of the epoch on which the effect of Gd III disappears requires non-LTE radiative transfer simulations.

\begin{figure}[t!]
\includegraphics[width = \columnwidth]{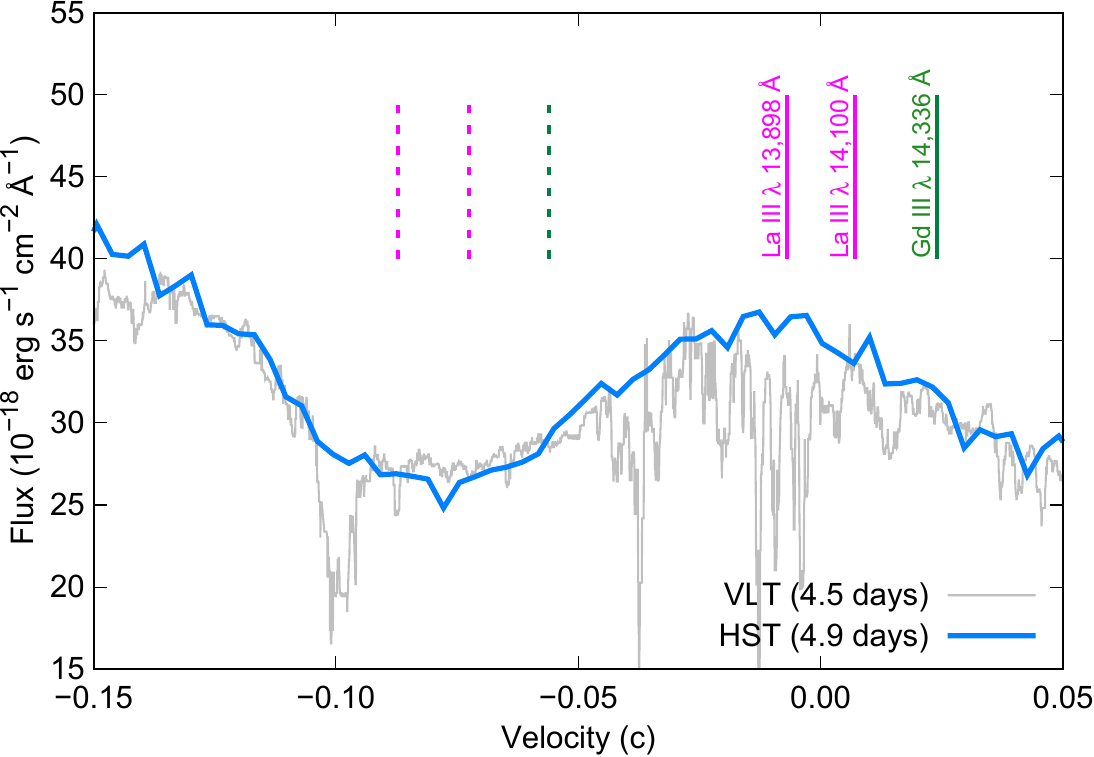} \\
\includegraphics[width = \columnwidth]{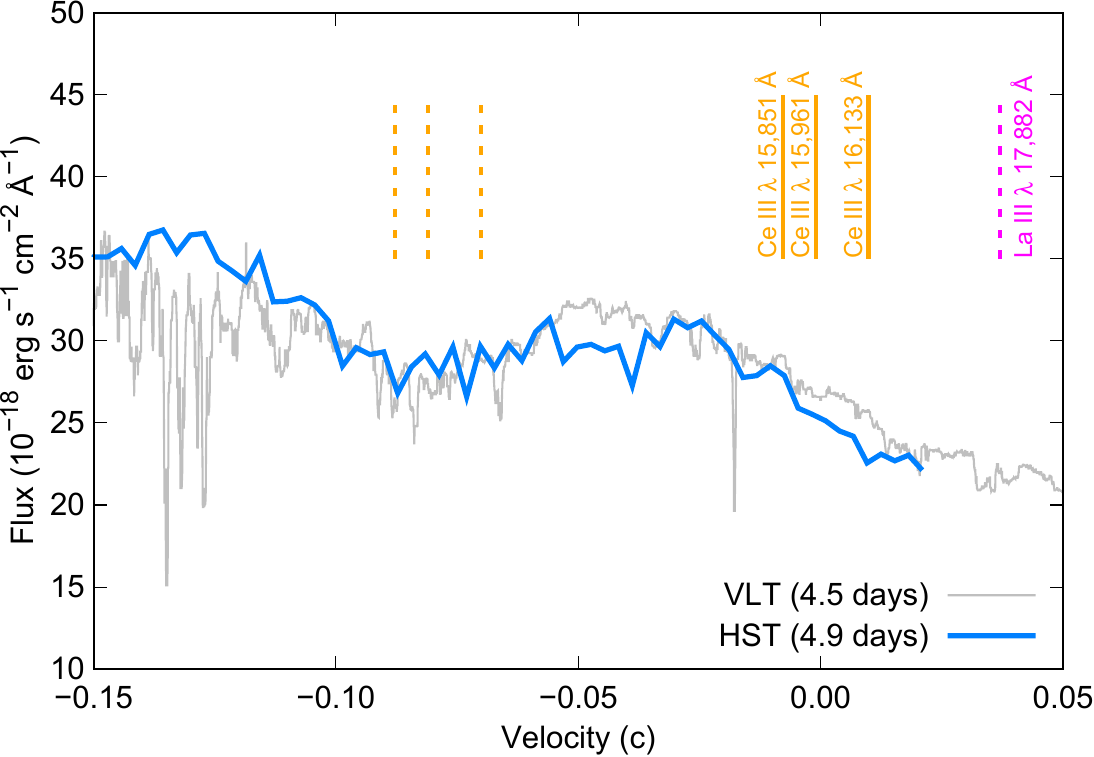}
\caption{Top: La III/Gd III line profile taken by the HST at 4.9 days \citep{tanvir2017emergence} and the VLT at 4.5 days \citep{pian2017spectroscopic, smartt2017kilonova} in blue and gray, respectively. Velocity offset is calculated assuming a rest wavelength of \(14,000\,\text{\AA}\). Transitions causing the features are highlighted in their rest and blueshifted wavelengths in solid and dashed lines, respectively. The blueshift is set as \(v=0.08\,c\). Bottom: same as top but for Ce III feature, with a rest wavelength of \(15,997\,\text{\AA}\).
\label{fig:feature}}
\end{figure}

Note that a thorough investigation of the La III/Gd III feature for AT2017gfo was challenging. Most spectra were taken with the Very Large Telescope (VLT) X-Shooter and the bottom panel of Figure \ref{fig:rad_1} shows that the feature coincides with the telluric absorption region. The only high-quality observation is the Hubble Space Telescope (HST) spectrum taken at 4.9 days, whose line profile is shown in the top panel of Figure \ref{fig:feature}. For comparison, the Ce III feature at \(4.5-4.9\) days is shown in the bottom panel.
The velocity is measured by assuming the rest wavelength of \(14,000\,\text{\AA}\) and \(15,977\,\text{\AA}\) for La III/Gd III and Ce III, respectively. Important lines in play are highlighted for both features at rest (solid) and blueshifted (dashed) wavelengths.
The shift observed from \(v=0\) in the emission peak of the Ce III feature is most likely to be explained by La III \(\lambda\,17,883\,\text{\AA}\) absorption, also shown in the bottom panel of Figure \ref{fig:feature}. La III/Gd III feature, overall, looks more prominent than that of Ce III, suggesting that multiple elements might be in effect. However, due to numerous factors influencing both features, it is still difficult to draw firm conclusions from a single-epoch spectrum. Future kilonovae observations should focus on following the time evolution of the spectra with space-based telescopes such as the HST or the James Webb Space Telescope.

Other Gd III transitions do not significantly affect the spectra due to their relatively low \(gf\)-values (Table \ref{tab:gd}). These transitions are also not detected in the chemically peculiar star HR 465 NIR spectrum, except for \(17,479\,\text{\AA}\) which showed a relatively weak line (right panels of Figure \ref{fig:pstar}). This supports the conclusion that the \(14,336\,\text{\AA}\) line is the only significant Gd III transition for kilonova NIR spectra. Furthermore, lines of Gd III with empirically calculated \(gf\)-values are distributed throughout the spectra, rather than being clustered like Ce III strong lines (see Figure \ref{fig:tau_cegd}). This minimizes their effect on the synthesized spectrum. Nevertheless, we have found that the theoretical average \(gf\)-value largely underestimates the true strength of transitions. The effect of all Gd III lines should be considered for a more thorough investigation, and further efforts should focus on providing a complete and accurate line list of Gd III.

In our study, we focused primarily on lanthanides, emphasizing the importance of Gd III, Ce III, and La III. While our analysis in Section \ref{sec:cand} also identifies actinides as potential candidates for NIR spectral investigations, we could not confirm their importance due to the lack of \(gf\)-values for individual transitions. Some theoretical calculations have aimed to provide complete line lists for actinides \citep{fontes2023, pognan2024, deprince2024}, but such calculations require the calibration of energy levels with experimental data, which is beyond the scope of this paper. Nevertheless, a comprehensive study of the importance of actinides remains essential for a thorough spectral investigation of kilonovae, and we leave such investigations for future work.

\section{Conclusions}\label{sec:conc}

To investigate \(r\)-process nucleosynthesis in neutron star mergers, it is important to identify element species likely to exhibit strong features in kilonova spectra.
In this paper, we used the energy levels available from different databases to construct transitions for all singly and doubly ionized elements, focusing mainly on allowed transitions occurring from low-lying energy levels. Our findings suggest that lanthanides and actinides are most likely to appear on the kilonova NIR spectrum. This is due to their low-lying energy levels that stem from their complex atomic structure,
involving the outer \(4f\), \(5d\), and \(6s\) orbitals for lanthanides, and the outer \(5f\), \(6d\), and \(7s\) orbitals for actinides

Among the candidate elements, we identified Gd III as the most significant, uninvestigated species due to its several unique properties. Unlike most other doubly ionized lanthanides, which have ground states involving only the \(4f\) orbital, the ground configuration of Gd III involves the \(5d\) orbital. This causes the element to have one of the lowest-lying energy levels, between which several transitions with relatively high \(gf\)-values occur. Notably, the \(14,336\,\text{\AA}\) transition was found to have a Sobolev optical depth comparable to that of Ce III lines. Furthermore, the \(14,336\,\text{\AA}\) line, along with the \(17,479\,\text{\AA}\) line, was also detected in the spectrum of HR 465, a chemically peculiar star known to have abundance patterns and ionization degrees similar to those observed in kilonovae \citep{tanaka2023cerium}. These three properties\(-\)unique atomic structure, relatively high \(gf\)-values, and emergence in chemically peculiar stars\(-\)make Gd III an interesting candidate for further kilonova spectral investigation.

Our radiative transfer simulations show that the Gd III \(14,336\,\text{\AA}\) line enhances the feature previously attributed to La III and causes its center to shift by about \(500\,\text{\AA}\). Since Gd III is expected to recombine earlier than La III, this effect is likely to be most prominent at early time. Therefore, monitoring the evolution of the \(\sim12,000\,\text{\AA}\) feature over time could help detect these changes.
Since this feature coincides with the telluric region, investigating the effect of Gd III and constraining its mass fraction in the ejecta remains challenging for AT2017gfo. This highlights the importance of space-based time-series observations of future kilonovae to thoroughly investigate this feature.

The elements most likely to explain the kilonova NIR spectral features are La III, Ce III, and Gd III.
Based on the currently available atomic data, the contribution of other elements with \(Z\leq88\) is expected to remain minimal. Meanwhile, no firm conclusions can be made about actinides (\(89\leq Z\leq 103\)) due to the lack of reliable transition data.
A thorough investigation requires complete and accurate line lists for all heavy elements. Therefore, further efforts to provide full experimentally calibrated energy levels and transition data for all species are necessary. The findings of this work should guide and prioritize such future studies.

\begin{acknowledgments}
We thank Nikolai Piskunov and Tanya Ryabchikova for their help and clarification on the Gd III data in the VALD database. We thank the anonymous referee for their constructive feedback. This work was supported by JST FOREST Program (Grant Number JPMJFR212Y, JPMJFR2136), NIFS Collaborative Research Program (NIFS22KIIF005, NIFS24KIIQ013), the Grant-in-Aid for Scientific Research from JSPS (21H04997, 23H00127, 23H04891, 23H04894, 23H05432, 24H00242), and the Grant-in-Aid for JSPS Fellows (22KJ0317). N.D. acknowledges support from Graduate Program on Physics for the Universe (GP-PU) at Tohoku University.
\end{acknowledgments}

\vspace{5mm}

\bibliography{kn_nir_gd}{}
\bibliographystyle{aasjournal}

\end{document}